\documentclass[runningheads]{llncs}

\usepackage[T1]{fontenc}
\def\doi#1{\href{https://doi.org/\detokenize{#1}}{\url{https://doi.org/\detokenize{#1}}}}
\usepackage{bbding}
\usepackage{tikz}
\usepackage{calc}
\usepackage{graphicx}
\usepackage{amsmath}
\usepackage{amsfonts}
\usepackage{listings}
\begin{document}
\title{VAFO-Loss: VAscular Feature Optimised Loss Function for Retinal Artery/Vein Segmentation}
\titlerunning{VAFO-Loss: VAscular Feature Optimised Loss Function}
%
\author{Yukun Zhou\inst{1,2,3}, Moucheng Xu\inst{1,2}, Yipeng Hu\inst{1,2,5}, Stefano B. Blumberg\inst{1,4}, An Zhao\inst{1,4}, Siegfried K. Wagner\inst{3}, Pearse A. Keane\inst{3}, and Daniel C. Alexander\inst{1,4}
}

%
\authorrunning{Y. Zhou et al.}
%
\institute{Centre for Medical Image Computing, University College London, London, UK \email{yukun.zhou.19@ucl.ac.uk}\\
\and
Department of Medical Physics and Biomedical Engineering, UCL, London, UK
\and
NIHR Biomedical Research Centre at Moorfields Eye Hospital NHS Foundation Trust and UCL Institute of Ophthalmology, London, UK
 \and 
Department of Computer Science, University College London, London, UK
\and
Wellcome/EPSRC Centre for Interventional and Surgical Sciences, London, UK
}
\maketitle         
\begin{abstract}
Estimating clinically-relevant vascular features following vessel segmentation is a standard pipeline for retinal vessel analysis, which provides potential ocular biomarkers for both ophthalmic disease and systemic disease. In this work, we integrate these clinical features into a novel vascular feature optimised loss function (VAFO-Loss), in order to regularise networks to produce segmentation maps, with which more accurate vascular features can be derived. Two common vascular features, vessel density and fractal dimension, are identified to be sensitive to intra-segment misclassification, which is a well-recognised problem in multi-class artery/vein segmentation particularly hindering the estimation of these vascular features. Thus we encode these two features into VAFO-Loss. We first show that incorporating our end-to-end VAFO-Loss in standard segmentation networks indeed improves vascular feature estimation, yielding quantitative improvement in stroke incidence prediction, a clinical downstream task. We also report a technically interesting finding that the trained segmentation network, albeit biased by the feature optimised loss VAFO-Loss, shows statistically significant improvement in segmentation metrics, compared to those trained with other state-of-the-art segmentation losses. \textcolor{red}{Code will be released with publication.}


\end{abstract}
%
%

\section{Introduction}

The significance of retinal vasculature for assessing ophthalmic disease has been well studied, such as venous beading serving as a reference for diagnosing diabetic retinopathy. Such retinal vascular features can further provide valuable insights into systemic disease \cite{wagner2020insights,cheung2021deep,wong2004hypertensive}, a field which has been termed `oculomics' \cite{wagner2020insights}. For example, increased artery tortuosity is associated with hypercholesterolaemia and hypertension \cite{cheung2011retinal}. Estimating these clinical features from segmentation maps is considered a standard pipeline in clinical research~\cite{cheung2011retinal,de2018clinically,seidelmann2016retinal}, as the ``intermediate'' segmentation algorithm maintains the pipeline's generalisability, transparency, and interpretability. Since manual vessel segmentation is tedious, there have been efforts to fully automate vessel segmentation.

\begin{figure}[t]
\centering
\includegraphics[width=0.9\columnwidth]{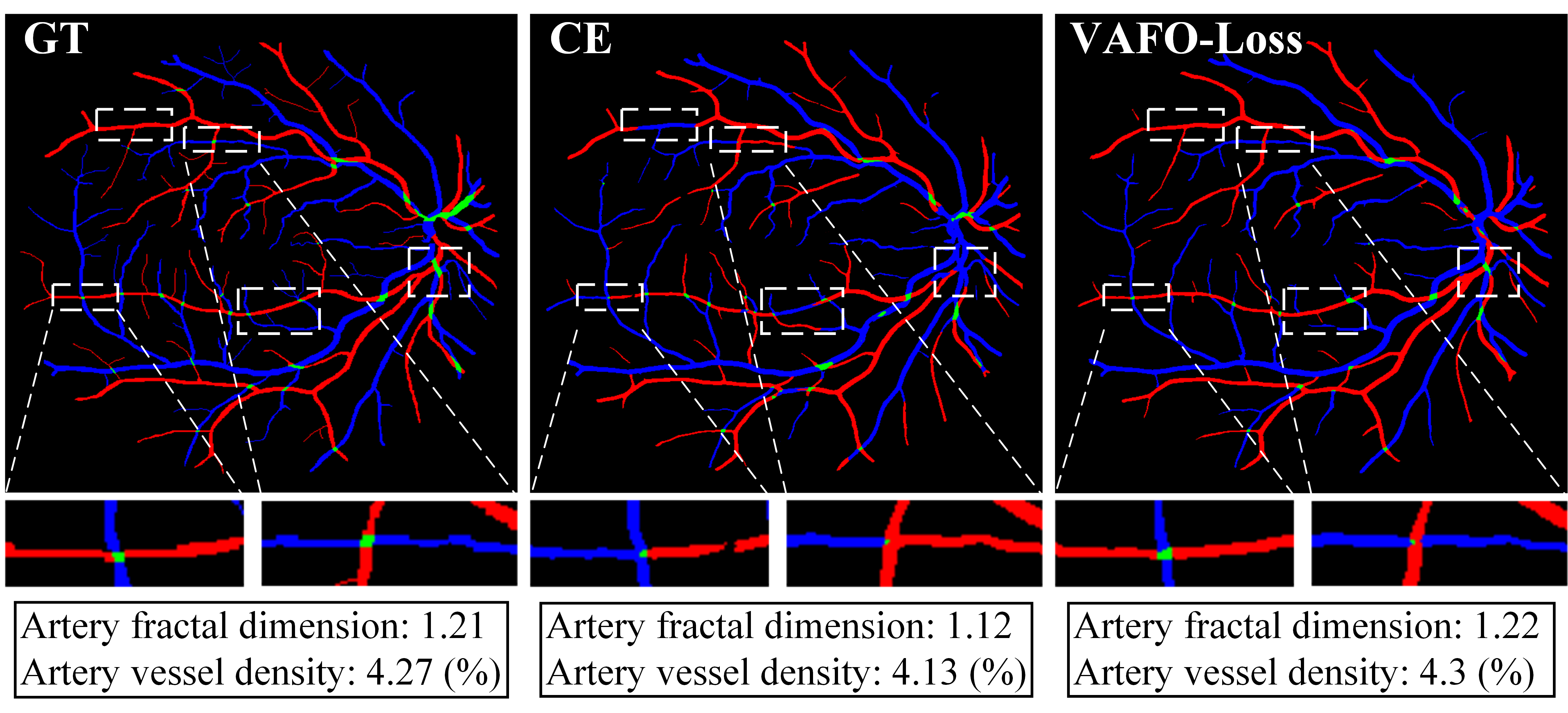} 

\caption{Examples of intra-segment misclassification, highlighted in white dash boxes. GT: manually labelled ground-truth, CE: segmentation with cross entropy loss, and VAFO-Loss: segmentation with the proposed VAFO-Loss. Red, blue, and green indicate arteries, veins, and uncertain pixels at intersections, respectively. Vessel density and fractal dimension derived from VAFO-Loss correspond to GT better than CE.} 

\label{fig1}
\end{figure}

Two main categories of methods, namely feature-based \cite{huang2018artery,mirsharif2013automated} and graph-based \cite{dashtbozorg2013automatic,estrada2015retinal,srinidhi2019automated,xie2020classification,zhao2019retinal}, segment arteries and veins from the retinal fundus photographs respectively based on hand-crafted features and topological knowledge. Recently, the development of deep learning models has further boosted the artery/vein segmentation performance enabled by the powerful capability of representation learning \cite{hemelings2019artery,galdran2019uncertainty,10.1007/978-3-030-87193-2_46}. Despite the progress, some artery/vein segmentation challenges remain such as intra-segment misclassification, which usually occurs at branch intersections where arteries and veins show high similarity and intersect with each other, as shown in Figure 1. This misclassification flips the prediction of arteries and veins for an entire segment between the intersections, which can generate a considerable error in vascular feature estimation. Although some methodologies, e.g. information fusion \cite{10.1007/978-3-030-87193-2_46}, morphology ranking \cite{li2020joint,chen2021tw}, topological constraints \cite{hu2019topology,zheng2021graph,shit2021cldice,chen2021tw,chen2020tr}, can address this issue to a degree, their efficacy of improving the resulting feature estimation has not been investigated. To the best of our knowledge, little work focuses on regularising networks to produce segmentation maps that can derive precise vascular features. The accuracy of the feature estimation is arguably more important for the downstream clinical tasks, such as oculomics, than segmentation itself.

In this work, we propose a vascular feature optimised loss function (VAFO-Loss) to obtain a segmentation map which can derive accurate vascular features. Highly sensitive vascular features are acutely aware of intra-segment misclassification and show a large error when such misclassification occurs. Formulating these sensitive features as loss functions optimises deep learning models to produce segmentation that is consistent with correct vascular features. Here we focus on encoding vessel density and fractal dimension in the proposed VAFO-Loss and show how this improves their accuracy as well as downstream tasks in oculomics. However, the concept extends to other image-based vascular features.  

We summarise our contributions as follows. 1) To our knowledge, we are the first to integrate clinical vascular features into a deep learning loss function VAFO-Loss. VAFO-Loss minimises the differences between the network output and the ground-truth in both the segmentation and its derived vascular features, which is found effectively reducing intra-segment misclassification. 2) Besides artery/vein segmentation metrics, we evaluate the estimation accuracy of vascular features and their utility in a downstream task. We also report that, experimentally, segmentation metrics may fail to indicate accurate vascular features, further highlighting the importance of optimising vascular features directly. 3) We show VAFO-Loss's effectiveness in improving vascular features' accuracy as well as a downstream clinical task, compared with state-of-the-art losses. We also demonstrate that VAFO-Loss can produce a statistically significant improvement in segmentation metrics.

\begin{figure}[t]
\centering
\includegraphics[width=0.95\columnwidth]{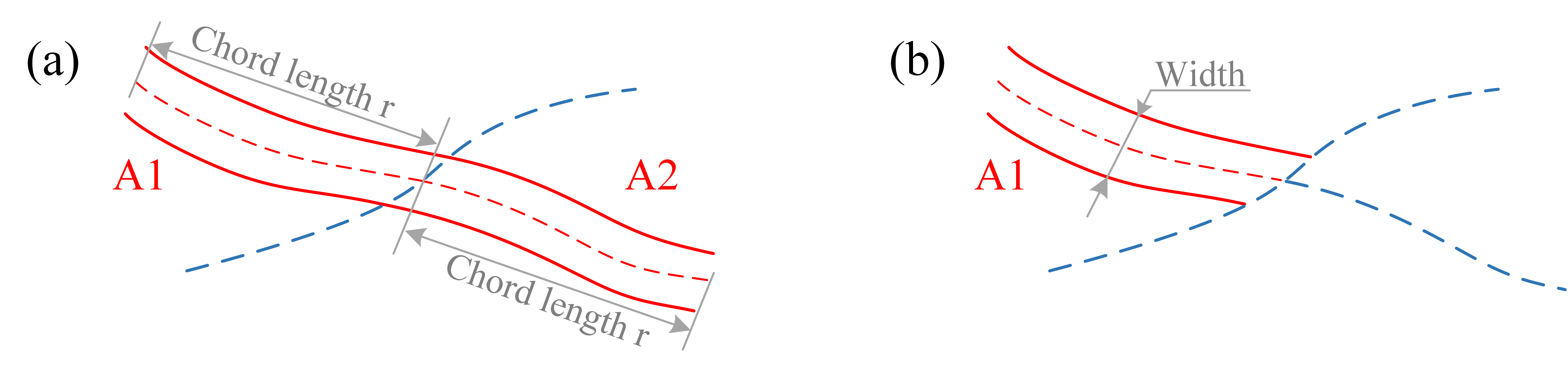} 
\caption{Schematic of a typical intra-segment misclassification. An artery segment $A2$ (red) in (a) is incorrectly classified as vein (blue) in (b). Dash lines indicate centrelines.} 

\label{fig2}
\end{figure}

\section{Methods}

We first analyse and compare how sensitive the three vascular features, tortuosity, vessel density, and fractal dimension, are with respect to intra-segment misclassification, using a relative error, and then select highly sensitive features, vessel density and fractal dimension, to construct VAFO-Loss.

\subsection{Relative Error from Intra-segment Misclassification}
In a typical intra-segment misclassification, a collection of arteries pixels $A$ is separated into two segments, $A1$ and $A2$, as shown in Figure \ref{fig2}(a). $A2$ is wrongly classified as veins in intra-segment misclassification, as shown in Figure \ref{fig2}(b). For illustrating the comparison of relative error, we set a simplified equal chord length of $r$ to $A1$ and $A2$, therefore $2r$ for $A$, and a uniform artery width. We use artery as an example in the following derivation. 

\subsubsection{Tortuosity.} Distance factor tortuosity $t(\cdot)$ calculates the ratio of centreline's arc length $arc(\cdot)$ to chord length $chord(\cdot)$~\cite{hart1999measurement}, e.g., $t(A)=\frac{arc(A)}{chord(A)}$. We can estimate the relative error by $\frac{\left|t(A)-t(A1)\right |}{t(A)}=\left | \frac{arc(A)}{chord(A)}-\frac{arc(A1)}{chord(A1)} \right |/\frac{arc(A)}{chord(A)}$. Since $chord(A1)=chord(A2)=r$ and $arc(A)=arc(A1)+arc(A2)$, we have

\begin{equation}\label{eq1}
\frac{\Delta t}{t(A)} =  \frac{\left |arc(A2)-arc(A1)\right |}{arc(A2)+arc(A1)}=\frac{\left |1-arc(A2)/arc(A1) \right |}{arc(A2)/arc(A1)+1}.
\end{equation}

\subsubsection{Vessel density.} Vessel density $v(\cdot)$ measures a ratio of vessel area $area(\cdot)$ to the whole area $v_{whole}$, e.g., $v(A)=\frac{area(A)}{v_{whole}}$. We estimate the relative error by $\left | \frac{area(A)}{v_{whole}}-\frac{area(A1)}{v_{whole}} \right |/\frac{area(A)}{v_{whole}}$. As $area(A)=arc(A)\times width(A)$ and width is uniform, relative error can be expressed as

\begin{equation}\label{eq2}
\frac{\Delta v}{v(A)}=\frac{\left |arc(A)-arc(A1) \right |}{arc(A)}=\frac{arc(A2)/arc(A1)}{arc(A2)/arc(A1)+1}.
\end{equation}

\subsubsection{Fractal dimension.} Fratal dimension $f(\cdot)$ evaluates vessel morphology complexity. Minkowski-Bouligand dimension (also known as box counting dimension) \cite{Falconer2004-vy} is used in this work. For artery $A$, the fractal dimension is $f(A)=\lim_{\varepsilon \rightarrow 0}\frac{\textrm{log}N_{A}(\varepsilon)}{\mathrm{log}(1/\varepsilon )}$ where $\varepsilon$ indicates square box size and $N_{A}(\varepsilon)$ represents the number of boxes required to cover artery A. In practice, we choose a set of box sizes $\varepsilon_C=\left \{2^i | i\in \mathbb{Z}, 2\leq 2^i\leq min\left\{h,w\right\} \right \}$ as $x$ axis, and $N_A(\varepsilon_C)$ as $y$ axis, where $h$ and $w$ indicate the image height and width, respectively. Least-square regression is used to fit a straight line with slope approximating $f(A)$. Denote error on $y$ axis $\Delta y=\left | \textrm{log}(N_A(\varepsilon_C))-\textrm{log}(N_{A1}(\varepsilon_C)) \right |=\textrm{log}\frac{N_A(\varepsilon_C)}{N_{A1}(\varepsilon_C)}$. When $\varepsilon_C$ is small, $N_A(\varepsilon_C)\approx \frac{area(A)}{\varepsilon_C^2}$, thus $\textrm{log}\frac{N_A(\varepsilon_C)}{N_{A1}(\varepsilon_C)}\approx \textrm{log}\frac{area(A)/\varepsilon_C^2}{area(A1)/\varepsilon_C^2}=\textrm{log}\frac{arc(A)}{arc(A1)}$, which shows $\Delta y$ is a constant determined by the arc ratio. This indicates that two lines (respectively for $A$ and $A1$) are approximately parallel, i.e. equal fractal dimensions. However, we identify that box counts $N_A(\varepsilon_C)$ is highly sensitive and therefore may be useful in reducing intra-segment misclassification. When $\varepsilon_C$ is small, the relative error of box counts is

\begin{equation}\label{eq3}
\frac{N_A(\varepsilon_C)-N_{A1}(\varepsilon_C)}{N_A(\varepsilon_C)}\approx\frac{area(A)-area(A1)}{area(A)}=\frac{arc(A2)/arc(A1)}{arc(A2)/arc(A1)+1}.
\end{equation}

\subsubsection{Relative error comparison.}

Combining Eq. \ref{eq1} and Eq. \ref{eq2}, vessel density shows larger relative error when $\frac{arc(A2)}{arc(A1)}>0.5$, which arguably holds at the intersections of large branches where the arc length of $A2$ and $A1$ shows limited difference. Additionally, Eq. \ref{eq3} has the same resultant form to that of Eq. \ref{eq2}. This indicates that vessel density and box counts approximate similar relative errors to this typical intra-segment misclassification. Alternatively, the varying box size of fractal dimension may be interpreted with the perspective of receptive fields, i.e., large boxes contribute to semantic information in large receptive fields, while small boxes contain high resolution information in local areas. This has partly motivated the formulation of vessel density and fractal dimension in the proposed VAFO-Loss for model training, summarised in section 2.2.

\subsection{VAFO-Loss Formulation}

Denote $S, T\in \left [0, 1  \right ]^{h\times w}$ as a segmentation map and a ground-truth map, respectively. We introduce the construction of vessel density loss $Loss_V(S,T)$ and box counts loss $Loss_B(S,T)$. The $Loss_V(S,T)$ can be expressed as  

\begin{equation}\label{eq5}
Loss_{V}(S,T)=\left | \frac{\sum S}{h\times w}-\frac{\sum T}{h\times w} \right |\leq \frac{\sum\left | S-T \right |}{h\times w},
\end{equation}

\noindent which interestingly reflects the fact that $Loss_V(S,T)$ approximates pixel-level segmentation loss function, the mean absolute error. Therefore, we use cross entropy loss function to replace $Loss_V(S,T)$, for its numerically preferred property in penalising misclassification for multi-class segmentation tasks.

For $Loss_B(S,T)$, we set the box sizes $\varepsilon=\left \{2^i | i\in \mathbb{Z}, 2\leq 2^i\leq min \left\{h,w\right\}\right \}$ and count the box number $N_S(\varepsilon)$ and $N_T(\varepsilon)$ and calculate $Loss_B(S,T)$ by

\begin{equation}\label{eq6}
Loss_B(S,T)=\frac{1}{\sqrt{\sum_{i}^{}\varepsilon_i^2}}\cdot \sum_{i}^{}\sqrt{\varepsilon_i\cdot  (\frac{N_T(\varepsilon_i)-N_S(\varepsilon_i)}{N_T(\varepsilon_i)})^2},
\end{equation}

\noindent where $\frac{N_T(\varepsilon)-N_S(\varepsilon)}{N_T(\varepsilon)}$ normalises multi-scale box errors to have the same magnitudes. $\varepsilon_i$ is empirically configured to weight more on the error of large-size box to regularise semantic information, to which intra-segment misclassification corresponds. $1/\sqrt{\sum_{i}^{}\varepsilon_i^2}$ scales $Loss_B(S,T)$ to the same level of magnitude to $Loss_V(S,T)$. For model training, $Loss_V(S,T)$ and $Loss_B(S,T)$ are combined with a weight $\lambda$ to form the VAFO-Loss as follows.

\begin{equation}\label{eq7}
Loss_{\textit{VAFO}}=Loss_V(S,T)+\lambda\cdot Loss_B(S,T).
\end{equation}

\section{Experiments}
\subsection{Experiment Setting}
We use public datasets, DRIVE-AV~\cite{staal2004ridge,hu2013automated}, LES-AV~\cite{orlando2018towards}, and HRF-AV~\cite{budai2013robust,hemelings2019artery}, to verify the performance of artery/vein segmentation and feature estimation. DRIVE-AV has 40 colour fundus photographs with size $(565, 584)$, where 20 images are for training and 20 for testing. LES-AV contains 22 images with a size of $(1620, 1444)$, 11 for training and 11 for testing. HRF-AV includes 45 images with size $(3504, 2336)$, 24 for training and 21 for testing. $10\%$ of training data are used for validation. \textbf{For our clinical downstream task}, we predict three-year ischaemic stroke incidence using derived retinal vascular features. Macula-centred retinal colour images of 1548 patients (774 with stroke and 774 controls) originate from the anonymous study (name and reference are removed for anonymity), a retrospective cohort data linkage project. A logistic regression model was trained with 60\% of the data to predict stroke incidence with input of a vascular feature, such as fractal dimension or vessel density, and test with 40\% data. Further demographic information is provided in Supplementary Materials Section 1. 

We use U-Net~\cite{ronneberger2015u} and BF-Net~\cite{10.1007/978-3-030-87193-2_46} as segmentation backbones. We substitute backbone's original loss respectively with VAFO-Loss and compared state-of-the-art losses (Active Contour (AC)~\cite{chen2019learning}, GraphCut (GC)~\cite{zheng2021graph}, clDice~\cite{shit2021cldice}), and compare their performance. All training hyperparameters were set the same as the previous work \cite{ronneberger2015u,10.1007/978-3-030-87193-2_46}. We use one Tesla T4 GPU (16GB) in all experiments. We halve BF-Net channels to include clDice which is computationally expensive and change the final activation function to softmax. The segmentation output is a class probability map including four categories, background, artery, vein, and uncertain pixel. All training images are resized to $(720, 720)$ to fit computation resources. For testing, the segmentation map is resized back to the original size for calculating metrics. For loss weight $\lambda$, we set it as 0.5 based on validation performance, whilst test data remains unseen in model development.

For evaluation, we measure the agreement of estimated vascular features to ground-truth features derived from manual artery/vein annotation. Following previous work \cite{cheung2021deep}, intra-class correlation coefficient (ICC) is used to evaluate the feature agreement. For the clinical evaluation, we use area-under-curve receiver operating characteristic (AUC-ROC) and area-under-curve precision-recall (AUC-PR) to evaluate the binary classification performance of the logistic regression model. We also calculate a weighted performance of artery, vein, and uncertain pixels \cite{10.1007/978-3-030-87193-2_46}, using segmentation metrics, F1-score (i.e. Dice), MSE, and IOU, together with topology metric, the Betti number error~\cite{hu2019topology,shit2021cldice}. \textit{p-values} from the Mann–Whitney U test are reported, when statistical comparisons are made. We implement the code of VAFO-Loss with Pytorch 1.9.

\begin{table}[t]
\caption{Segmentation results with backbone BF-Net on DRIVE-AV, LES-AV, and HRF-AV. Betti error evaluates the topological correctness of segmentation maps. $\mathrm{ICC_A}$ evaluate the agreement of artery fractal dimension to that derived from ground-truth maps. \textit{p-value} of Mann–Whitney U test between VAFO-Loss and clDice is reported, as clDice is the most competitive loss function among others.}
\label{table1}
\centering
\renewcommand\arraystretch{1.0}
{
\smallskip\begin{tabular}{llllll}
\hline
                   \multicolumn{6}{c}{\textbf{DRIVE-AV}}\\ \hline
Loss               & F1-score $\uparrow$       & IOU $\uparrow$    & MSE $\downarrow$     & Betti error$\downarrow$  & $\mathrm{ICC_{A}}(95\%CI)$$\uparrow$  \\ \hline

AC~\cite{chen2019learning}      & 67.31$\pm$2.08    & 52.22$\pm$1.97  & 3.31$\pm$0.32    &   14.04$\pm$3.91 &  0.64(0.09\text{-}0.86) \\
GC~\cite{zheng2021graph}           &68.87$\pm$2.48     & 52.57$\pm$2.63   & 3.29$\pm$0.35       &    13.96$\pm$4.69 &   0.71(0.27\text{-}0.89)  \\ 
clDice  \cite{shit2021cldice}           & 70.27$\pm$1.1    & 54.55$\pm$1.36   & 3.22$\pm$0.18       &   11.48$\pm$1.54 &   0.74(0.45\text{-}0.9)   \\
VAFO-Loss  &  \textbf{73.04$\pm$0.58}      &   \textbf{57.99$\pm$0.7}  &  \textbf{2.93$\pm$0.06}      &    \textbf{7.75$\pm$1.21}  &  $\mathbf{0.85(0.62\text{-}0.94)}$ \\

\textit{p-value}               &  1.36e-3       &  1.36e-3   &   3.87e-3    &   5.07e-3 & N/A \\ \hline

\multicolumn{6}{c}{\textbf{LES-AV}}\\ \hline
Loss               & F1-score $\uparrow$       & IOU $\uparrow$    & MSE $\downarrow$     & Betti error$\downarrow$  & $\mathrm{ICC_{A}}(95\%CI)$$\uparrow$  \\ \hline

AC~\cite{chen2019learning}      & 66.15$\pm$2.45    & 51.09$\pm$2.8  & 2.63$\pm$0.27    &   9.3$\pm$2.86  &  0.68(0.3\text{-}0.92) \\
GC~\cite{zheng2021graph}           &65.88$\pm$2.58     & 50.18$\pm$2.82   & 2.68$\pm$0.18       &   6.76$\pm$2.58  & 0.83(0.56\text{-}0.95)   \\ 
clDice  \cite{shit2021cldice}           & 67.2$\pm$2.23    & 51.87$\pm$2.3   & 2.5$\pm$0.19       &   4.51$\pm$0.69   &   0.82(0.47\text{-}0.96) \\
VAFO-Loss  &  \textbf{69.87$\pm$1.56}      &   \textbf{54.98$\pm$1.61}  &  \textbf{2.32$\pm$0.1}      &  \textbf{3.04$\pm$0.66}  &  $\mathbf{0.86(0.51\text{-}0.96)}$  \\

\textit{p-value}               &  1.95e-3      &   9.39e-4  &  5.38e-3    &  3.06e-3  & N/A  \\ \hline

\multicolumn{6}{c}{\textbf{HRF-AV}}\\ \hline
Loss               & F1-score $\uparrow$       & IOU $\uparrow$    & MSE $\downarrow$     & Betti error$\downarrow$  & $\mathrm{ICC_{A}}(95\%CI)$$\uparrow$  \\ \hline

AC~\cite{chen2019learning}      & 68.22$\pm$0.8    & 53.09$\pm$0.73  & 2.26$\pm$0.04    &  10.81$\pm$3.59  & 0.86(0.68\text{-}0.95)  \\
GC~\cite{zheng2021graph}           &68.67$\pm$1.06     & 53.77$\pm$1.28   & 2.15$\pm$0.05       &  12.93$\pm$4.32  & 0.83(0.59\text{-}0.93)   \\ 
clDice  \cite{shit2021cldice}           & 68.83$\pm$0.45    & 54.04$\pm$0.57   & 2.13$\pm$0.05       &   9.48$\pm$1.1  &  0.89(0.74\text{-}0.96)  \\
VAFO-Loss  &  \textbf{71.19$\pm$0.58}      &   \textbf{56.48$\pm$0.73}  &  \textbf{1.96$\pm$0.03}      &   \textbf{6.82$\pm$1.16}  &  $\mathbf{0.92(0.8\text{-}0.97)}$  \\

\textit{p-value}               &   9.39e-4     &  9.39e-4   &    9.39e-4  &   2.76e-3    &   N/A \\ \hline

\end{tabular}}
\end{table}

\subsection{Experiment Results}

\textbf{Vascular feature estimation performance.} Table \ref{table1} shows the feature estimation performance with backbones BF-Net (results with U-Net are consistent and can be found in Supplementary Materials Section 2). VAFO-Loss increased $\mathrm{ICC_{A}}$ (artery fractal dimension) by $11\%$, $4\%$, and $3\%$ compared to clDice. This verifies that our VAFO-Loss indeed improved the vascular features' accuracy. Visualisation results are shown in Figure \ref{fig1}.

\begin{figure}[t]
\centering
\includegraphics[width=0.95\columnwidth]{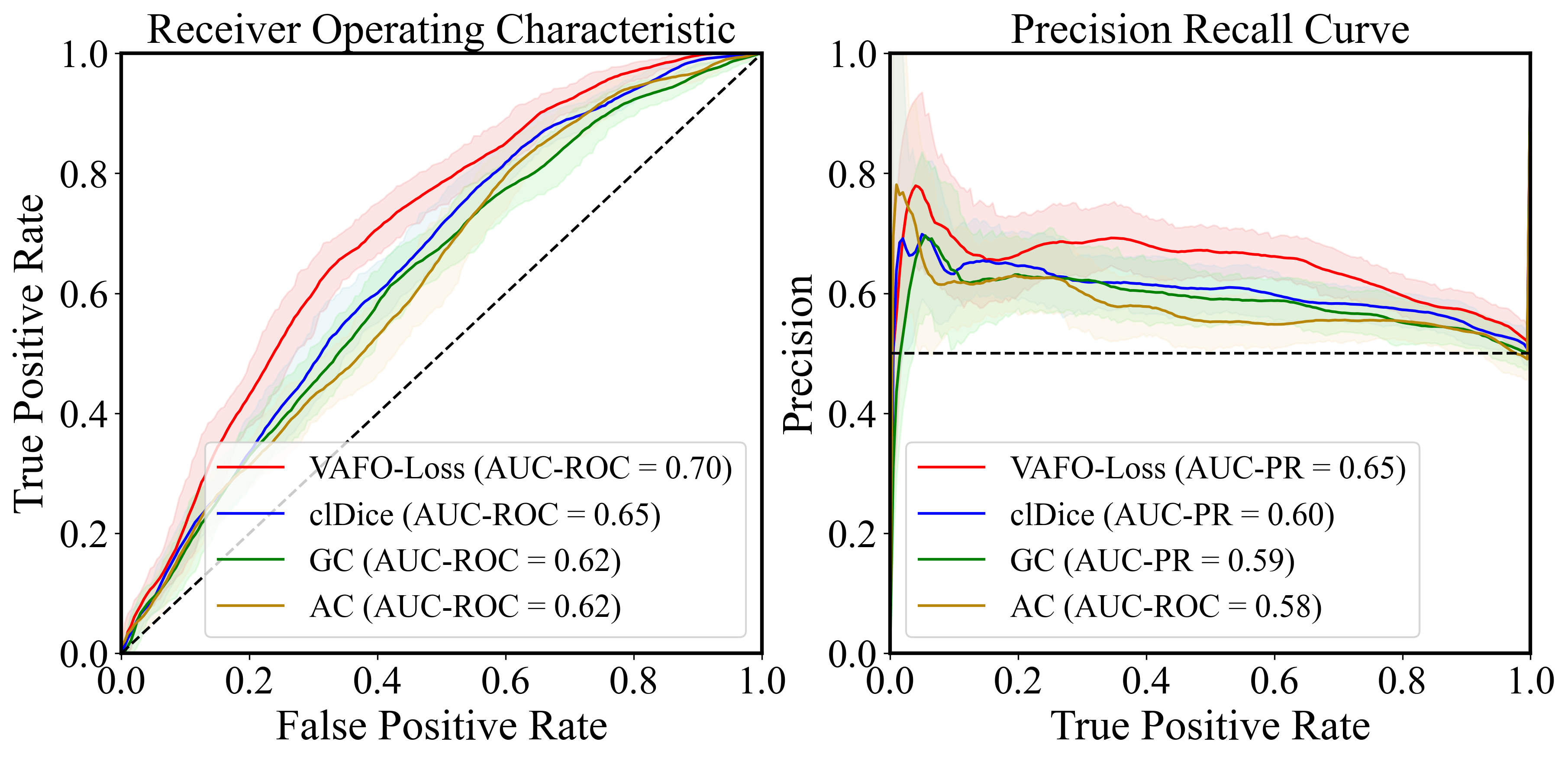} 

\caption{ROC and PR curves for predicting stroke incidence with artery fractal dimension from different segmentation loss functions. AUC-ROC and AUC-PR are listed in legends. Bootstrapped confidence intervals, $[5^{th},95^{th}]$ percentiles of AUC-ROC and AUC-PR, are plotted in corresponding colour shades.} 

\label{fig3}
\end{figure}

\textbf{Quantitative clinical impact.} We use vascular features to predict three-year ischaemic stroke incidence. The results with the input of artery fractal dimension are shown in Figure \ref{fig3}. AUC-ROC achieved 0.7 and AUC-PR achieved 0.65 with VAFO-Loss, outperforming those of the other three compared loss functions. This also demonstrates that VAFO-Loss improved downstream task performance with accurate vascular features, at varying different cut-off values (Results with other features are in Supplementary Materials Section 3).

\textbf{Segmentation metrics.} Table \ref{table1} also includes segmentation results with backbones BF-Net. VAFO-Loss achieved best performance in all segmentation metrics (e.g., F1-score increased by $2.75\%$, $2.67\%$, and $2.36\%$ in DRIVE-AV, LES-AV, and HRF-AV), as well as topological correctness (Betti error decreased by $3.73\%$, $1.47\%$, and $2.66\%$). All \textit{p-values} were smaller than 5.07e-3, showing statistically significant improvement over the compared loss functions.

\begin{figure}[t]
\centering
\includegraphics[width=0.95\columnwidth]{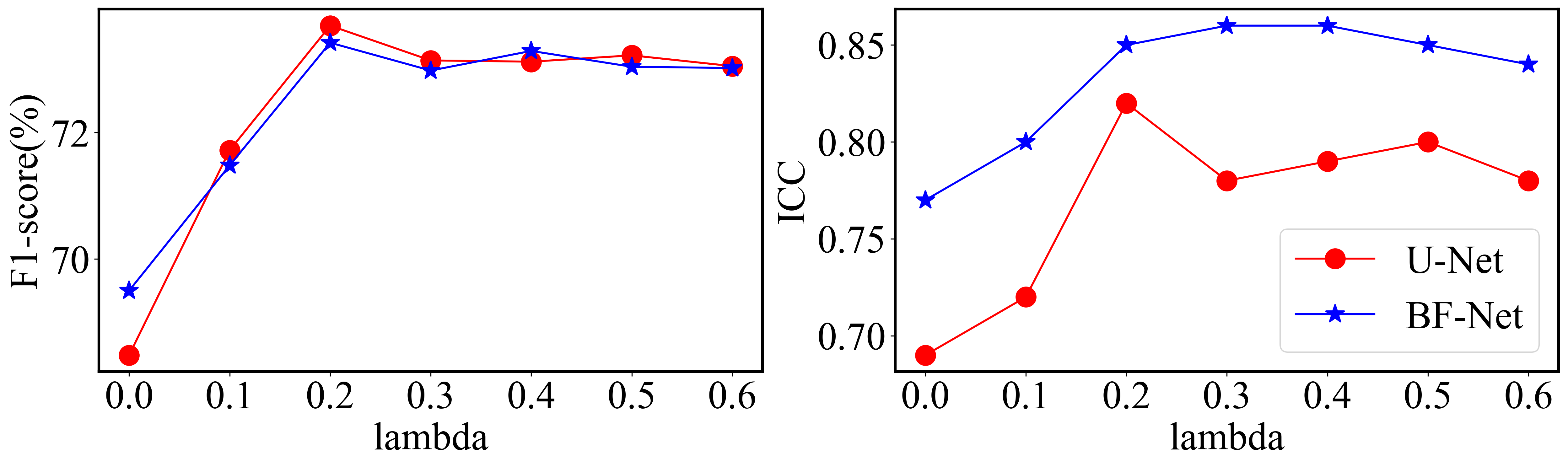} 
\caption{Segmentation performance with different $\lambda$ values for the VAFO-Loss. Left: F1-score, and right: the agreement of artery fractal dimension.} 

\label{fig4}
\end{figure}

\textbf{Impact of the hyperparameter $\lambda$.} We investigate the impact of different loss weights $\lambda$. Figure \ref{fig4} depicts the performance on DRIVE-AV dataset, with backbone U-Net (red line) and BF-Net (blue line). When $\lambda=0$, it reduces to cross entropy loss. The introduction of VAFO-Loss ($\lambda=0.1$ and $\lambda=0.2$) rapidly increases F1-score and fractal dimension agreement. After $\lambda$ is larger than 0.2, the performance stays relatively stable. This verifies that segmentation performance is not sensitive to the loss weight $\lambda$.

\textbf{Computation efficiency.} VAFO-Loss requires $O(n^2)$ computational complexity for a two-dimensional image with a size of $n\times n$, which is at the same level of complexity as AC, GC, and clDice. The proposed VAFO-Loss on average took 1.64s for a batch size of 2 and image size of $(720,720)$, while AC, GC, and clDice respectively took 1.87s, 1.91s, and 2.17s. The inference time is the same.

\section{Discussion and Conclusion}

In this paper, we introduce a new loss function that directly encodes computation of downstream vascular features used as clinical disease markers, to optimise segmentation networks. Experimental results concluded that not only does the loss improve the estimation of vascular features themselves, but indirectly forces the network to produce better segmentation than those with current state-of-the-art loss functions. Moreover, we demonstrate that the derived vascular features provided better disease biomarkers, in an example application of stroke prediction. From Figure \ref{fig3}, we observe that U-Net is comparable to BF-Net in F1-score while has worse performance in fractal dimension, which suggests that segmentation metrics can fail to represent the estimation accuracy of vascular features. This further highlights the benefits of optimising features directly, as proposed in this work.

Future work will focus on exploring encoding additional features into feature optimised loss functions. Although we show improvements in the accuracy of some features such as vessel density and fractal dimension, intra-segment misclassifications still arise in local areas with complicated vasculature, that can disrupt other important global vascular features. Additionally, we believe that embedding physicians' knowledge of vessel growth and bifurcation in deep learning models may further benefit artery/vein segmentation. Feature optimised loss function contributes to downstream clinical research, such as oculomics, and potentially promotes the deployment of automated AI techniques in clinical application. Further future work will evaluate VAFO-Loss on large-scale clinical datasets and a wider range of disease diagnosis tasks.
 
%
%


%
%
%
\bibliographystyle{splncs04}
\bibliography{reference}

\section{Demographic of clinical dataset (Table 1)}

\begin{table}[ht]
\caption{Demographic of clinical dataset. Age is a risk factor of incident stroke, while vascular features are complementary potential biomarkers. Please note that neither age nor gender is used in the logistic regression model. We compare the stroke prediction performance using only derived vascular features.}
\label{tables1}
\centering
\begin{tabular}{l|c|c|c}
 & With stroke & Control & \textit{p-value} \\ \hline
Age average (standard deviation)      &     74.44(11.45)       &         63.33(13.14)  &   $<0.001$   \\ \hline
Female (\% of total population)   &       394(50.9\%)      &         402(51.93\%)   & 0.5398     \\ \hline
\end{tabular}
\end{table}

\section{Segmentation results with backbone U-Net (Table 2)}

\begin{table}[ht]
\caption{Segmentation results with backbone U-Net on DRIVE-AV, LES-AV, and HRF-AV. Betti error evaluates the topological correctness of segmentation maps. $\mathrm{ICC_A}$ evaluate the agreement of artery fractal dimension to that derived from ground-truth maps. \textit{p-value} of Mann–Whitney U test between VAFO-Loss and clDice is reported, as clDice is the most competitive loss function among others.}
\label{tables2}
\centering
\renewcommand\arraystretch{1.0}
{
\smallskip\begin{tabular}{llllll}
\hline
                   \multicolumn{6}{c}{\textbf{DRIVE-AV}}\\ \hline
Loss               & F1-score $\uparrow$       & IOU $\uparrow$    & MSE $\downarrow$     & Betti error$\downarrow$  & $\mathrm{ICC_{A}}(95\%CI)$$\uparrow$  \\ \hline

AC     &68.4$\pm$1.42     & 53.29$\pm$1.69   & 3.18$\pm$0.13    & 10.64$\pm$0.84   &  0.74(0.36\text{-}0.9) \\
GC         &69.23$\pm$2.05     & 53.03$\pm$2.37   & 3.27$\pm$0.22       &  23.22$\pm$2.34  &  0.77(0.44\text{-}0.91) \\ 
clDice           & 70.34$\pm$1.11    & 55.64$\pm$1.38   & 3.05$\pm$0.15       &  10.98$\pm$0.92  & 0.75(0.36\text{-}0.89) \\
CD-Loss  &  \textbf{73.22$\pm$0.98}      &   \textbf{58.25$\pm$1.21}  &  \textbf{2.85$\pm$0.1}      &  \textbf{7.92$\pm$1.02}   &   $\mathbf{0.8(0.52\text{-}0.92)}$ \\

\textit{p-value}              &  1.36e-3   &  5.38e-3   &  2.39e-2    &   1.36e-3  &  N/A \\ \hline

\multicolumn{6}{c}{\textbf{LES-AV}}\\ \hline
Loss               & F1-score $\uparrow$       & IOU $\uparrow$    & MSE $\downarrow$     & Betti error$\downarrow$  & $\mathrm{ICC_{A}}(95\%CI)$$\uparrow$  \\ \hline

AC      & 62.83$\pm$2.32    & 47.4$\pm$2.6  & 2.88$\pm$0.16    &  8.42$\pm$0.75  & 0.66(0.31\text{-}0.82)  \\
GC       & 63.69$\pm$1.78    & 47.99$\pm$1.91   & 2.83$\pm$0.14       & 10.69$\pm$2.84   &    0.67(0.12\text{-}0.89)      \\ 
clDice          &63.87$\pm$1.94     & 48.55$\pm$1.9   & 2.86$\pm$0.12       &  8.42$\pm$1.31 &  0.65(0.24\text{-}0.83) \\

CD-Loss  &  \textbf{65.93$\pm$1.32}      &   \textbf{50.66$\pm$1.51}  &  \textbf{2.61$\pm$0.25}      & \textbf{4.76$\pm$1.15}  &  $\mathbf{0.72(0.4\text{-}0.92)}$  \\

\textit{p-value}              &    2.17e-3     & 7.29e-3    &    2.15e-2  &   3.13e-3 & N/A \\ \hline

\multicolumn{6}{c}{\textbf{HRF-AV}}\\ \hline
Loss               & F1-score $\uparrow$       & IOU $\uparrow$    & MSE $\downarrow$     & Betti error$\downarrow$  & $\mathrm{ICC_{A}}(95\%CI)$$\uparrow$  \\ \hline

AC      & 69.93$\pm$0.98    & 55.11$\pm$0.88  & 2.11$\pm$0.05    &  9.31$\pm$0.48  &  0.87(0.69\text{-}0.95)  \\
GC       &70.48$\pm$0.63     & 55.73$\pm$0.82   &2.18$\pm$0.07      &12.84$\pm$2.4   &  0.88(0.8\text{-}0.95) \\ 
clDice          & 70.14$\pm$1.04    & 56.12$\pm$0.67   & 2.13$\pm$0.04       & 9.39$\pm$0.46   & 0.88(0.82\text{-}0.95)  \\
CD-Loss  &  \textbf{72.17$\pm$0.66}      &   \textbf{57.74$\pm$0.73}  &  \textbf{1.91$\pm$0.02}      &  \textbf{6.61$\pm$0.52}   & $\mathbf{0.91(0.79\text{-}0.97)}$  \\

\textit{p-value}              &   1.36e-3     & 1.36e-3    &    9.39e-4  &  3.67e-3  & N/A  \\ \hline

\end{tabular}}
\end{table}

\section{Logistic regression with other features (Fig.1, Fig.2)}

\begin{figure}[ht]
\centering
\includegraphics[width=0.8\columnwidth]{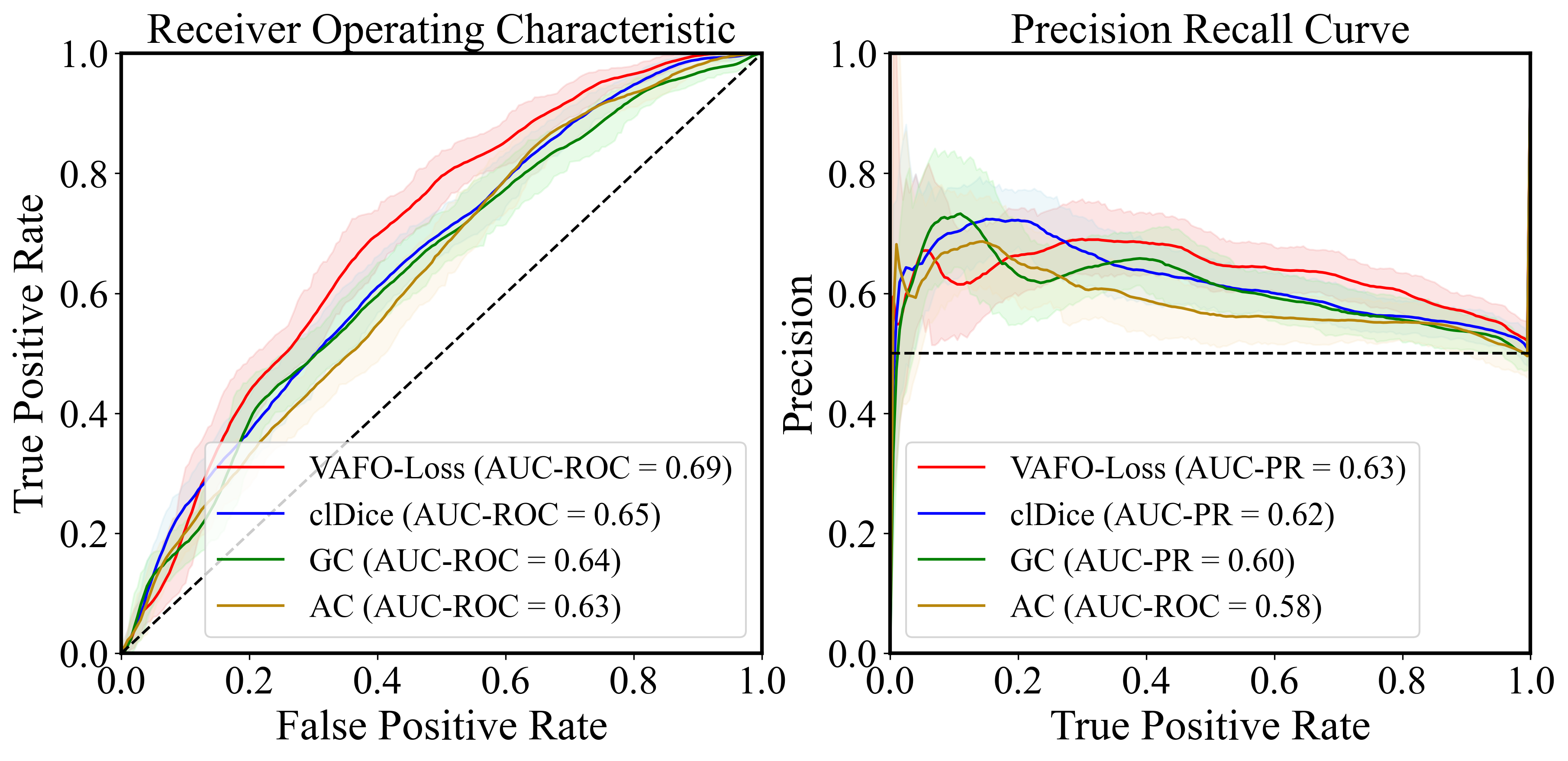} 

\caption{ROC and PR curves for predicting stroke incidence with \textbf{artery vessel density} from different segmentation loss functions. AUC-ROC and AUC-PR are listed in legends. Bootstrapped confidence intervals, $[5^{th},95^{th}]$ percentiles of AUC-ROC and AUC-PR, are plotted in corresponding colour shades.} 

\label{figs1}
\end{figure}

\begin{figure}[ht]
\centering
\includegraphics[width=0.8\columnwidth]{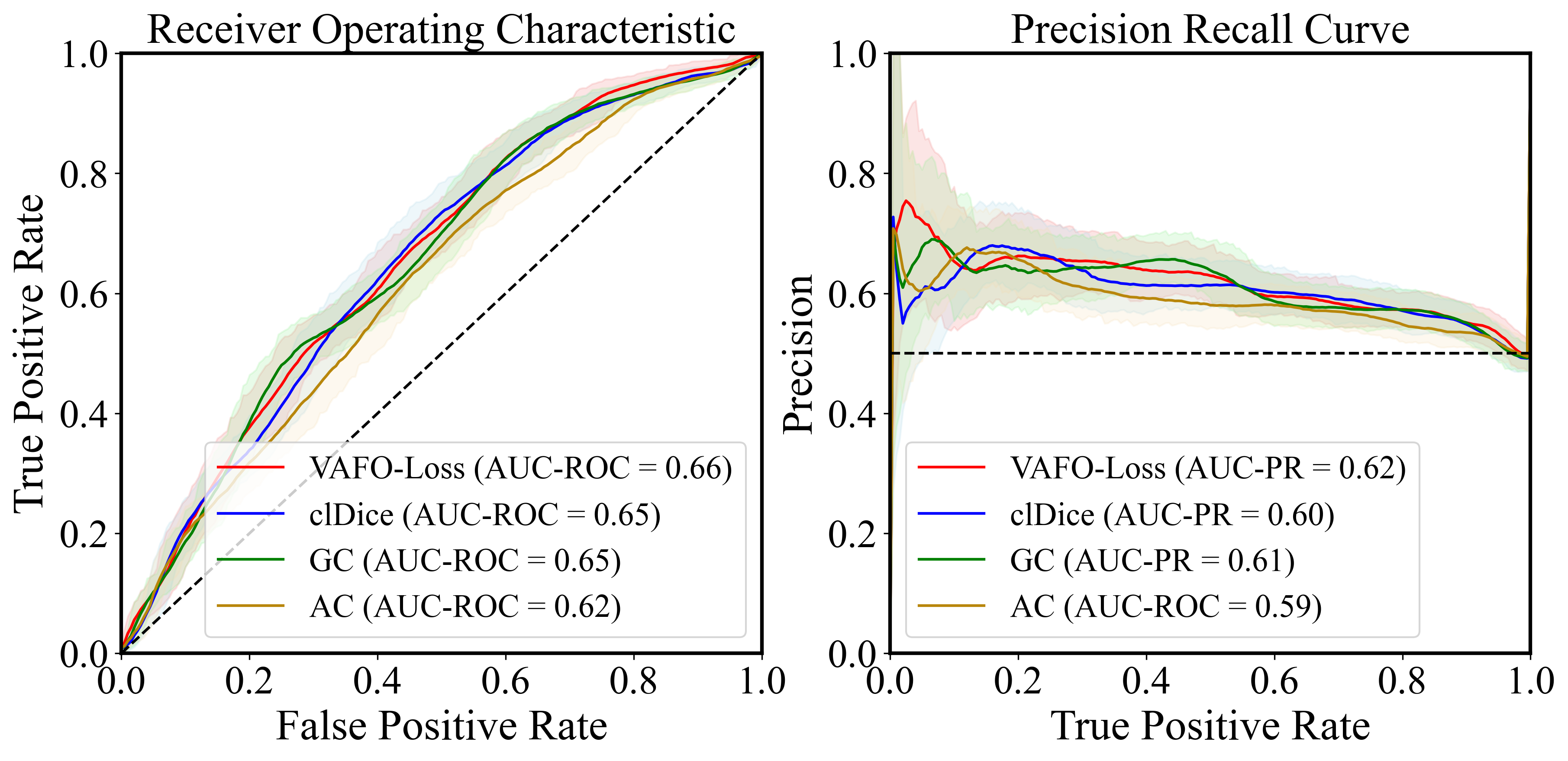} 

\caption{ROC and PR curves for predicting stroke incidence with \textbf{vein fractal dimension} from different segmentation loss functions. AUC-ROC and AUC-PR are listed in legends. Bootstrapped confidence intervals, $[5^{th},95^{th}]$ percentiles of AUC-ROC and AUC-PR, are plotted in corresponding colour shades.} 

\label{figs2}
\end{figure}

\end{document}